\documentclass[10pt, two column, twoside]{IEEEtran}
\usepackage{color}
\usepackage[colorlinks, linkcolor=color1, anchorcolor=blue, citecolor=color1]{hyperref}

\usepackage{amssymb}
\usepackage{amsmath}
\usepackage[lined,boxed,commentsnumbered, ruled]{algorithm2e}
\usepackage{mathrsfs}
\usepackage{algorithmic}
\usepackage{bm}
\usepackage{hhline}

\usepackage{pgfplots}
\usepackage{tikz}
\usetikzlibrary{arrows}
\usepackage{subfigure}
\usepackage{graphicx,booktabs,multirow}

\definecolor{colorhkust}{RGB}{20,43,140}
\definecolor{colortsinghua}{RGB}{116,52,129}
\definecolor{color1}{RGB}{128,0,0}

\date{}

\begin{document}

        \title{Task-Oriented Communications for 6G: Vision, Principles, and Technologies}
\author{Yuanming Shi, Yong Zhou, Dingzhu Wen, Youlong Wu, Chunxiao Jiang, and Khaled B. Letaief
        \thanks{Y. Shi, Y. Zhou, D. Wen, and Y. Wu are with ShanghaiTech University. C. Jiang is with Tsinghua University. K. B. Letaief is with The Hong Kong University of Science and Technology.}
                }

\maketitle

\maketitle

\begin{abstract}
Driven by the interplay among artificial intelligence, digital twin, and wireless networks, 6G is envisaged to go beyond data-centric services to provide intelligent and immersive experiences. To efficiently support intelligent tasks with customized service requirements, it becomes critical to develop novel information compression and transmission technologies, which typically involve coupled sensing, communication, and computation processes. 
To this end, task-oriented communication stands out as a disruptive technology for 6G system design by exploiting the task-specific information structures and folding the communication goals into the design of task-level transmission strategies. 
In this article, by developing task-oriented information extraction and network resource orchestration strategies, we demonstrate the effectiveness of task-oriented communication principles for typical intelligent tasks, including federated learning, edge inference, and semantic communication. 
\end{abstract}

\section{Introduction}

By integrating the communication, sensing, and computation capabilities into the space-air-ground integrated networks, 6G is envisaged to support five typical usage scenarios, including hyper connectivity and coverage, real-time broadband communications, extremely ultra-reliable and low-latency communications, integrated sensing and communication, as well as ubiquitous intelligence services, thereby fulfilling the vision of ``connected intelligence" and ``digital twin" \cite{Yuanming_JSAC22}. 
This can be achieved by enabling the functionalities of data sensing, processing, transmission, and decision making at the network edge to deliver scalable, trustworthy, reliable, and customized intelligent services, including edge training and edge inference for artificial intelligence (AI) models. 
Specifically, by keeping private raw data at the local devices
and only exchanging model parameters, wireless federated learning (FL) is able to embed the model training process at the network edge, while preserving data privacy and reducing communication overhead \cite{chen2021distributed}. 
By splitting deep neural networks (DNN) between edge devices and servers and leveraging the sensing and computation capabilities of edge devices, edge model inference is able to deliver real-time, private, and secure intelligent services \cite{zhuo2019edge}. 
It thus has been envisioned that 6G will go beyond the mobile internet with data-oriented services (e.g., voice, text, image, and video transmission) to support intelligent tasks and services (e.g., AI model training and inference), thereby enabling wide applications ranging from auto-driving to smart cities \cite{Yang_IEEENetwork22}.

\begin{figure*}[t] \centering
	\includegraphics[scale = 0.47]{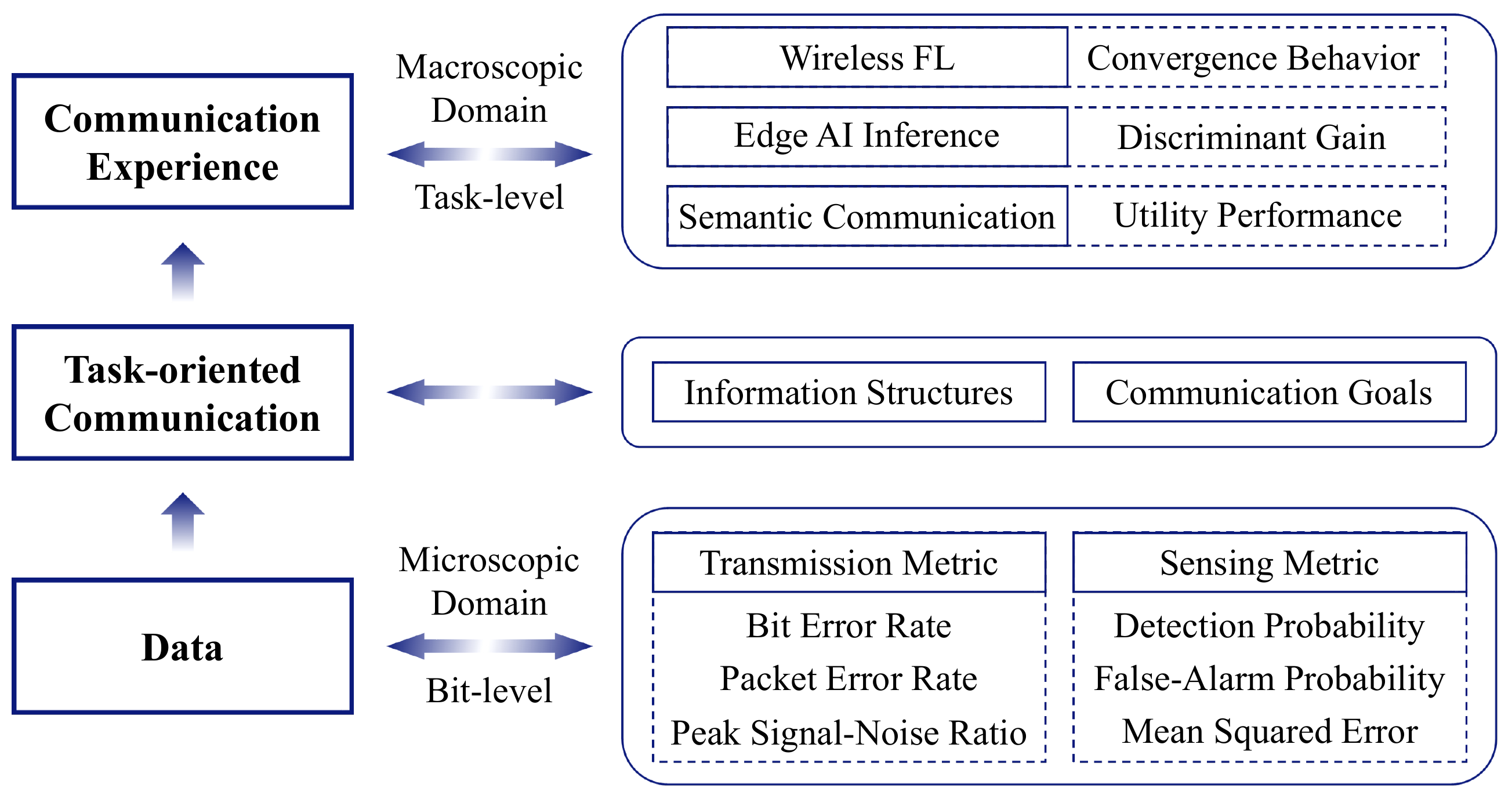}
	\caption{Task-oriented communications: a paradigm shift from ``bit-level transmission" to ``task-level transmission".}
	\label{Fig:0}
\end{figure*}

\begin{figure*}
        \centering
        \subfigure[Wireless FL for smart healthcare]{
         \begin{minipage}[b]{0.33\textwidth}
          \centering
          \includegraphics[width=1\textwidth]{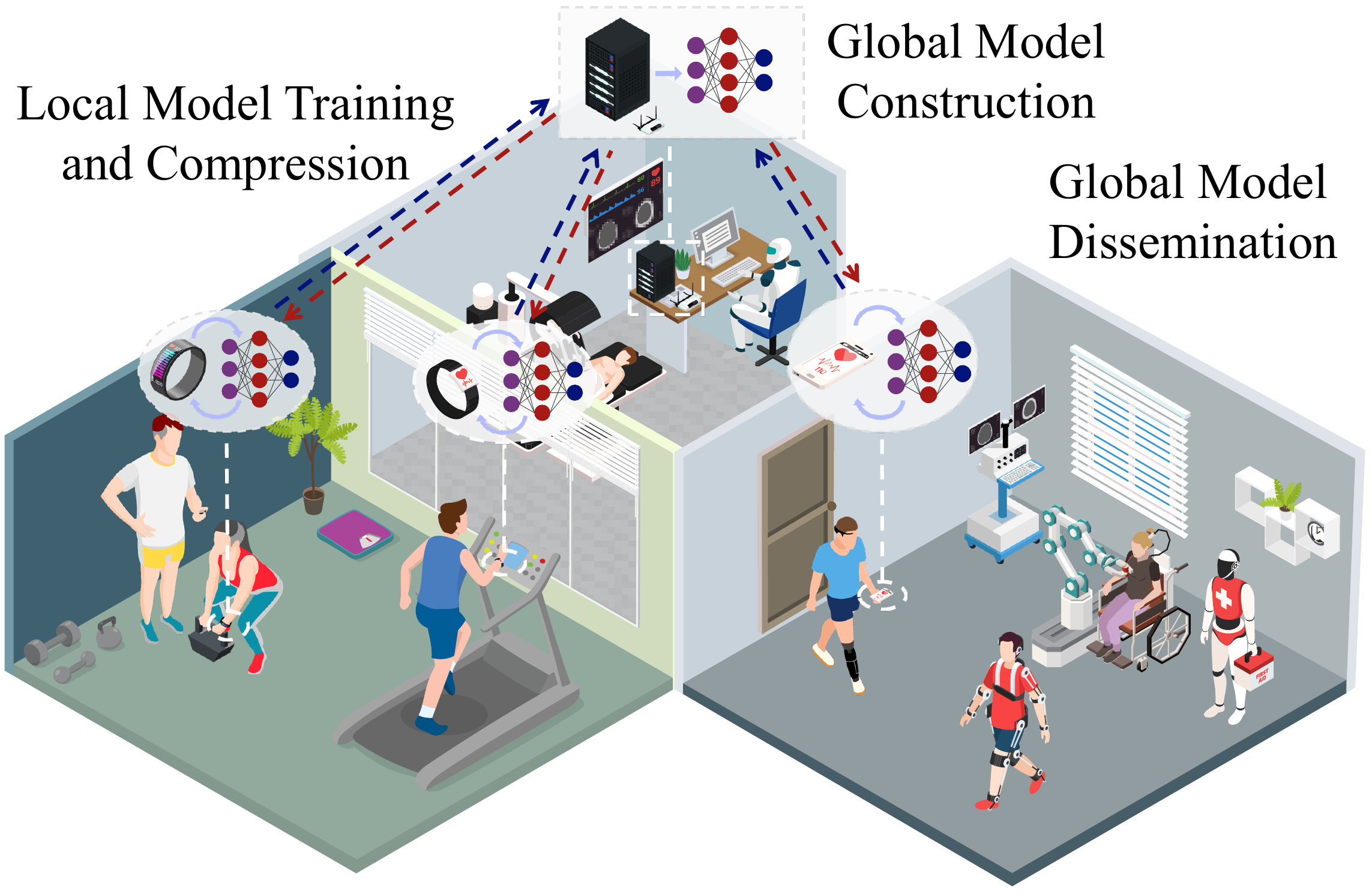}
         \end{minipage}
        }
        \subfigure[Edge AI inference for industrial IoT]{
         \begin{minipage}[b]{0.27\textwidth}
          \centering
          \includegraphics[width=1\textwidth]{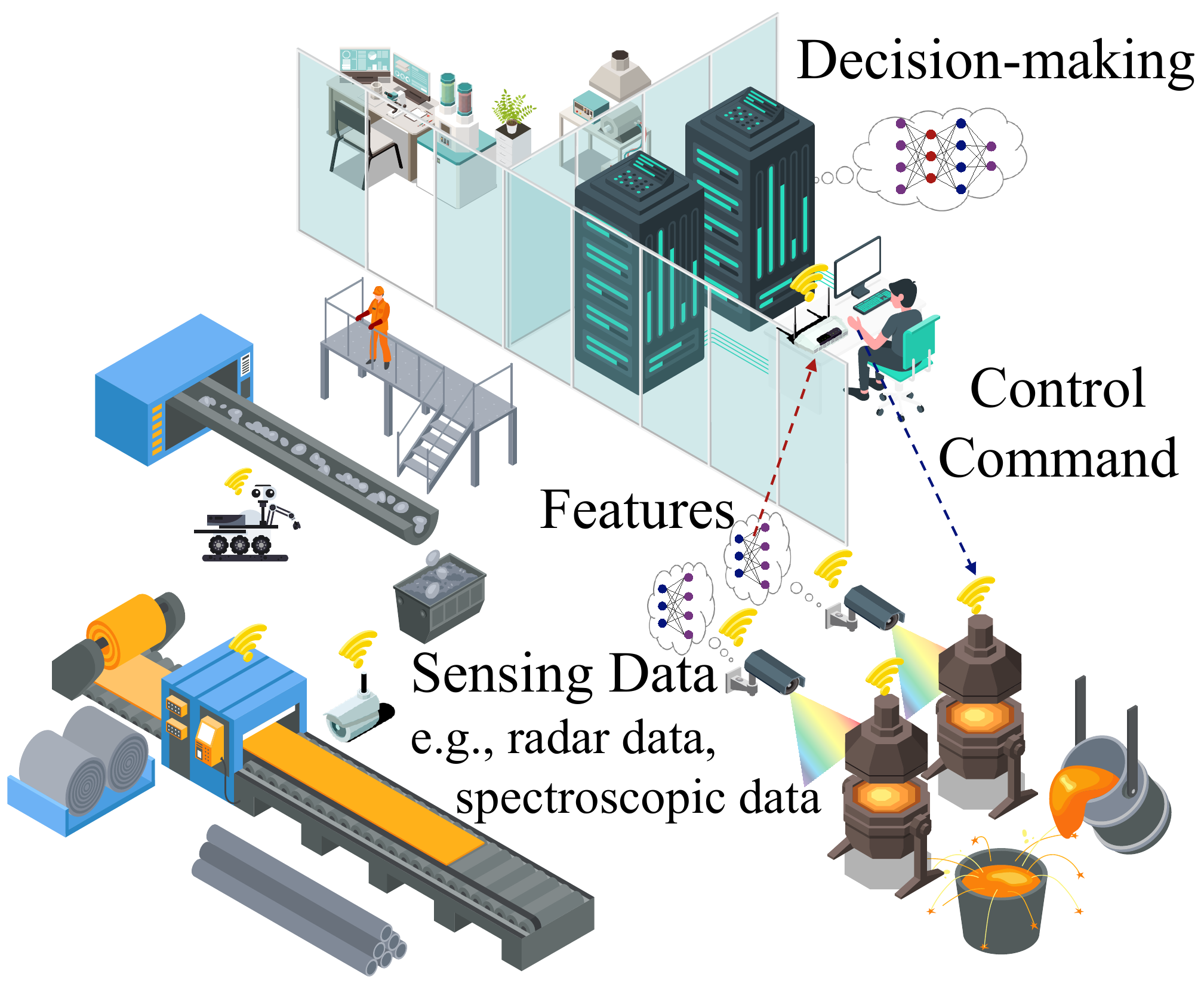}
         \end{minipage}
        }
        \subfigure[Semantic communication for autonomous driving]{
         \begin{minipage}[b]{0.34\textwidth}
          \centering
          \includegraphics[width=1\textwidth]{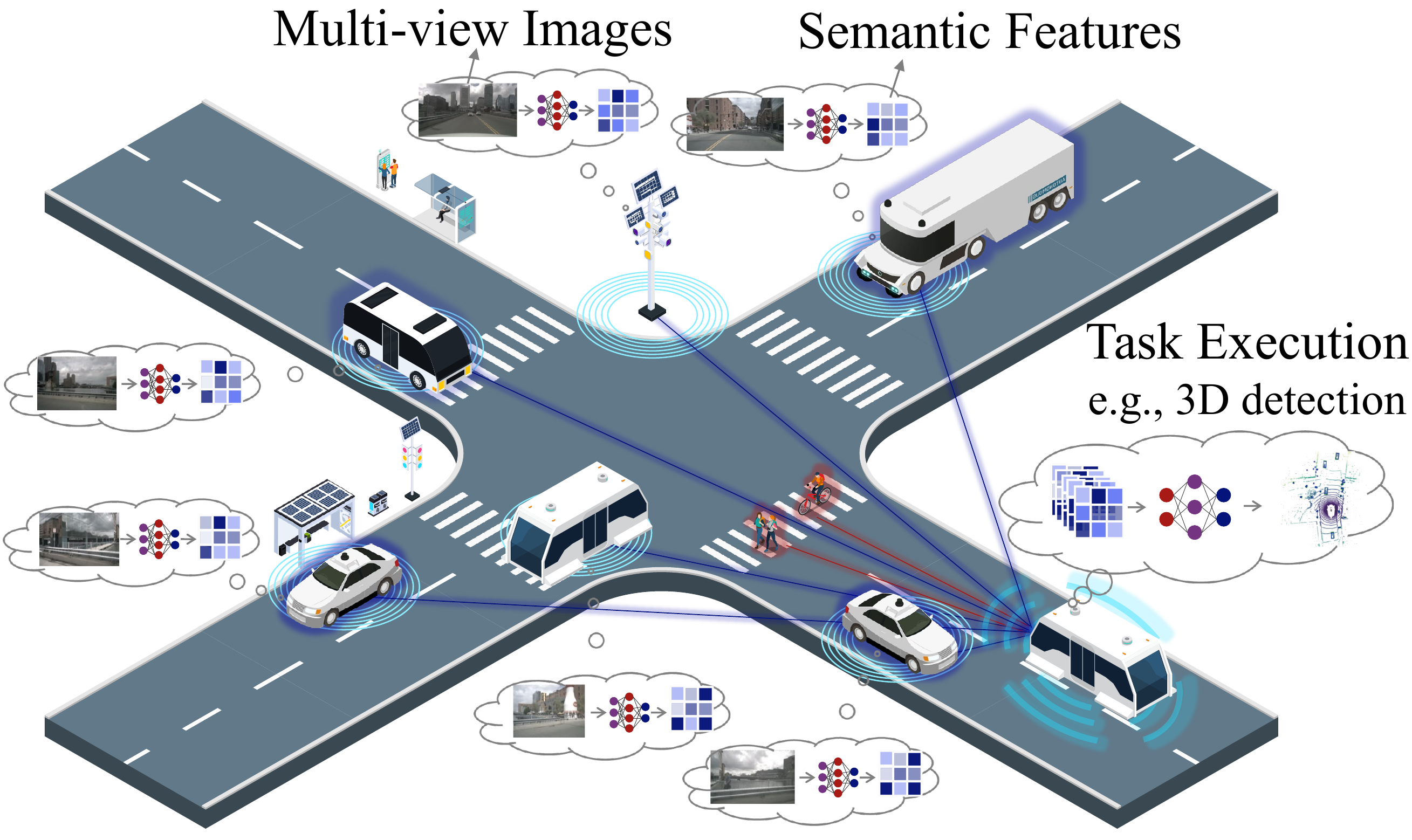}
         \end{minipage}
        }
        \caption{Typical applications of task-oriented communications.} 
       \label{fig_label}
\end{figure*}

The efficient delivery of diversified and trustworthy intelligent services imposes unique challenges for communication strategies design in wireless networks with limited network resources and customized service requirements. Specifically, the quality of intelligent services is normally characterized by the efficiency of completing specific tasks (e.g., convergence performance of edge training tasks), which cannot be solely evaluated by conventional task-agnostic performance metrics, such as throughput, delay, and reliability. It thus becomes infeasible to provide a unified performance metric for task-oriented services, for which the goal of communications needs to be folded into the system design. Besides, the task-relevant information is basically sufficient to support downstream intelligent decision making. 
Hence, extracting and transmitting only task-relevant information significantly reduces the end-to-end latency, which however violates the optimality condition of Shannon’s separate source-channel coding for the long block-length transmission of bit sequences. Furthermore, an intelligent task typically involves multiple complicated processes, which behave similarly to biological intelligence systems, including sensing for data acquisition (environmental perceptron), communication for data transmission (activation signal transportation), and computation for making intelligent decisions (brain). To support real-time intelligent services with limited network and hardware resources, it is crucial to reveal the interplay among sensing, communication, and computation for network resource optimization \cite{xu2022edge}. 

To address the challenges for delivering intelligent services, task-oriented communication provides a new paradigm to support ``connected intelligence" in 6G by exploiting the information structures \cite{lu2015structural} and folding the communication goals \cite{gunduz2022beyond} into the 6G system design. 
Task-oriented communication will thus go beyond transmitting bits with bit-level performance metrics in the microscopic domain (e.g., bit/packet error rate) to focus on communication experiences with task-level performance metrics in the macroscopic domain (e.g., learning rate and inference accuracy). 
Specifically, instead of sending the whole data and ignoring the information structures, task-oriented communication can reduce the communication load by delivering only task-relevant information, e.g., gradient sparsification for edge training and feature extraction for edge inference. 
By embedding the goals of communications into the system design, task-oriented communication can significantly improve the communication efficiency by developing goal-oriented end-to-end communication strategies, e.g., model aggregation over multiple access channels and semantic-level transmission via joint source and channel coding. By further characterizing the sensitivity and importance of the transmitted information for intelligent services, task-oriented communication can provide a holistic way for network resource orchestration across sensing, communication and computation, e.g., importance-aware data/feature sampling and transmission for edge training and inference tasks. 

In this article, we advocate a novel paradigm of task-oriented communications that goes beyond data-oriented services to support intelligent services by designing task-specific data sensing, signal transmission, and resource allocation strategies with the assistance of underlying information structures and communication goals, as shown in Fig. \ref{Fig:0}.
We discuss the vision, motivations, and principles of task-oriented communications to support scalable wireless FL, high-quality edge inference, and communication-efficient semantic communication, with typical applications in Fig. \ref{fig_label}. 
The main contributions of this article are three-fold.
Specifically, to support scalable wireless FL, we develop ``learning task"-oriented model compression and model aggregation strategies by exploiting the spatio-temporal model correlations and the signal superposition structure of wireless channels \cite{yang2020over}. 
To provide high-quality inference services, we propose ``inference task"-oriented feature compression and transmission schemes by utilizing the information bottleneck (IB) mechanism to characterize the rate-distortion tradeoff and by integrating the sensing, communication, and computation processes \cite{wen2022task}. 
To enable communication-efficient semantic communication, we provide a general information-theoretic framework to extract and compress task-relevant information, and present how to apply joint source-channel coding (JSCC) for achieving high performance  while being compatible with digital communication systems \cite{Wen_arXiv22}. 
We summarize the key features of these typical applications in Table I.

\begin{table*}[]
\centering
\caption{Key features of wireless FL, edge AI inference, and semantic communication.}
\begin{tabular}{|c|c|c|c|}
\hline
                      & Enabling technologies                                                                                                   & Performance metrics                                                                                          & Typical applications                                                                                 \\ \hline
Wireless FL           & \begin{tabular}[c]{@{}c@{}}Model compression \\ Over-the-air computation \\ Device scheduling\end{tabular}             & \begin{tabular}[c]{@{}c@{}}Convergence speed \\ Test accuracy\end{tabular}                 & \begin{tabular}[c]{@{}c@{}}Smart healthcare \\ Financial industry \\ Internet of energy\end{tabular} \\ \hline
Edge AI inference      &     \begin{tabular}[c]{@{}c@{}} Feature extraction and compression \\ Integration of sensing, communication, and computation \\ Network resource management\end{tabular}                                                                                                                      &  \begin{tabular}[c]{@{}c@{}}Inference accuracy \\ Inference latency\end{tabular}                                                                                                              &     \begin{tabular}[c]{@{}c@{}}   Self-service supermarkets \\  Virtual assistants \\  Intelligent agriculture\end{tabular}    \\ \hline
Semantic communication & \begin{tabular}[c]{@{}c@{}}Joint source-channel coding \\ AI technologies (e.g., Transformer)   \end{tabular} & \begin{tabular}[c]{@{}c@{}}Semantic error\\ Perception distortion \\ Classification distortion\end{tabular} & \begin{tabular}[c]{@{}c@{}}Autonomous driving \\ Telehealth \\ Environmental monitoring 
\end{tabular} \\ \hline
\end{tabular}
\end{table*}

\section{Task-Oriented Communications for Wireless Federated Learning}
This section presents ``learning task"-oriented model compression and aggregation to enable scalable wireless FL. 

\begin{figure*}[t] \centering
\includegraphics[scale = 0.48]{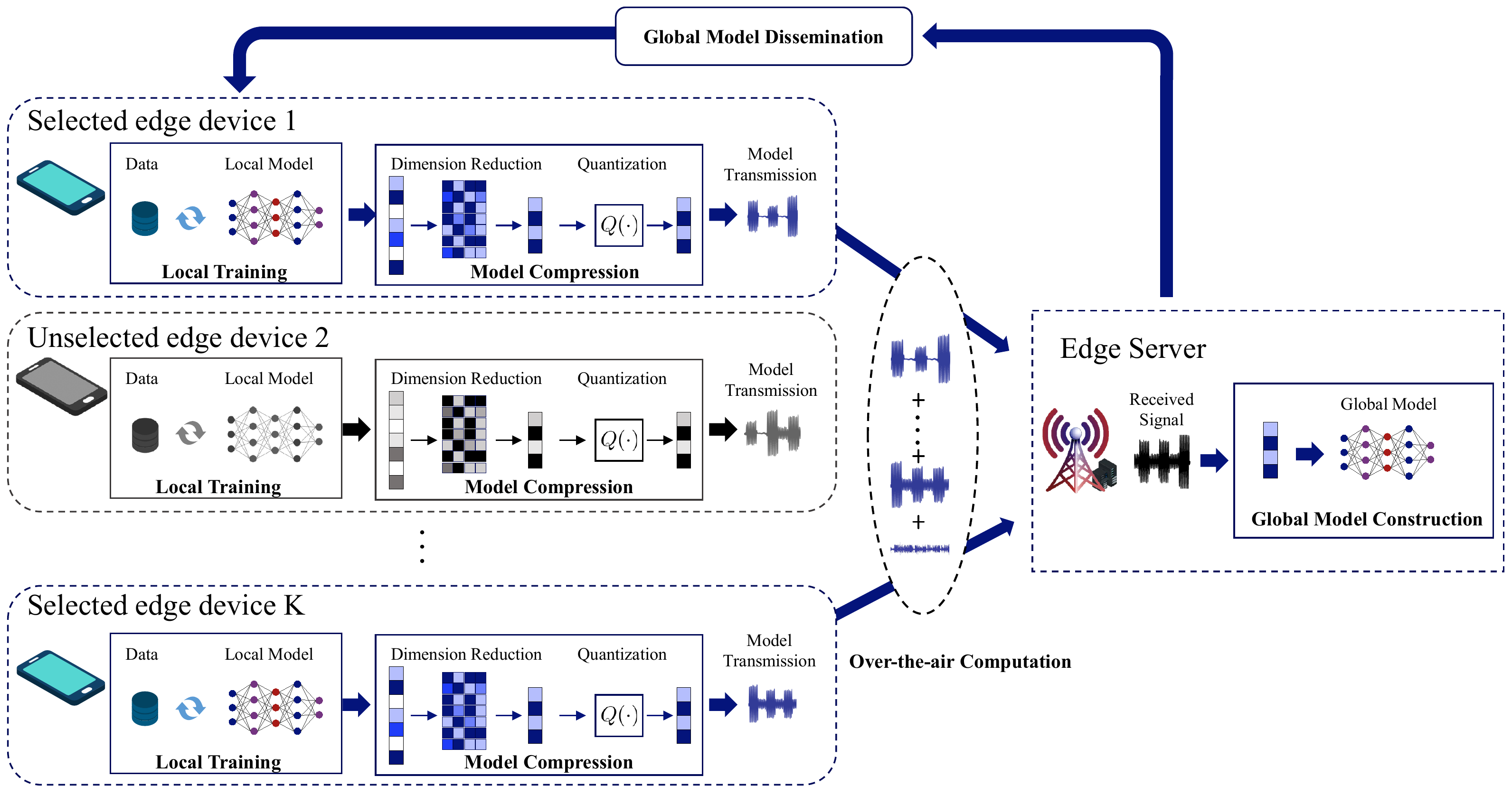}
\caption{An illustration of task-oriented model compression and model aggregation for FL over wireless networks.}
 \label{Fig:1}
\end{figure*}

\subsection{Vision}

As an emerging privacy-preserving distributed learning framework, wireless FL enables multiple edge devices to collaboratively train a global statistical model, wherein only model parameters are exchanged, as shown in Fig. \ref{Fig:1}.
The training performance of FL highly depends on the amount and quality of data available at each edge device, which further rely on the strategy design and resource allocation for data sensing.
After the data sensing process, FL generally consists of three steps, i.e., global model dissemination, local model training, and global model aggregation. 
Consisting of local model compression, transmission, and construction, the global model aggregation is a critical performance-limiting step to enable scalable FL \cite{Vincent_SPM22}. 
Particularly, massive devices transmit high-dimensional model parameters to the edge server over capacity-limited wireless links across multiple learning rounds, leading to severe communication bottleneck. 
However, the conventional source coding and data transmission methods are agnostic to FL tasks and fail to exploit the characteristics of locally trained models, wireless propagation channels, and learning procedure behaviors to facilitate the communication strategies design. To this end, task-oriented model compression and model aggregation are presented in this section. 

\subsection{Task-Oriented Model Compression}

The compression of local models, which can be viewed as a distributed lossy source coding problem, is critical to reduce the communication cost. As the conventional compression methods typically aim to reduce the amount of information that needs to be transmitted while guaranteeing some distortion measure, regardless of the underlying task, they fail to leverage the unique properties of FL tasks, including the global model aggregation structure, learning algorithm behavior, and spatio-temporal correlations among local models.
To address this issue, we develop task-oriented model compression methods (i.e., sparsification and quantization) by exploiting the information structures of local models to reduce the communication load while achieving a high learning performance. 

\subsubsection{Sparsification} 
Since each local model has an intrinsic sparsity structure and different entries of local models have diverse impacts on the learning performance, task-oriented sparsification can nullify unimportant model entries to effectively reduce the communication load without degrading the learning performance.
%By leveraging the intrinsic sparsity structure of local models, sparsification can reduce the number of delivered model parameters by nullifying unimportant model parameters, which have a vanishing impact on the model's accuracy. 
The sparsified model parameters can be further projected into low-dimensional signals for uplink transmission, from which the aggregated global model can be reconstructed by using the compressed sensing method. 
Another unique feature of FL is that the local models are temporally correlated in consecutive learning rounds as they vary slowly over time and also spatially correlated across different edge devices as the local datasets are sampled from the same global distribution. 
Both temporal and spatial correlations can be integrated with sparsification to enhance the learning performance from different perspectives, e.g., reducing the communication load and assisting the design of model aggregation schemes. 
%For instance, the server can substitute unsampled entries of a sparsified model with the corresponding entries in the previous iteration to improve compression efficiency, while each local model can be multiplied with a scaling factor based on the level of spatial correlation to enhance learning performance.

\subsubsection{Quantization} 
Sparsification can be utilized together with quantization to further reduce the communication load by adopting either scalar quantization (i.e., separately map each model element into a discrete quantity) or vector quantization (i.e., jointly map a set of model parameters into a finite-bit quantity). 
Similarly, we can also leverage the model correlation among different edge devices and across different learning rounds to improve quantization efficiency. 
Inspired by Wyner-Ziv and Berger–Tung coding, to avoid any probabilistic assumptions on the model and temporal correlations for designing lossy distributed source coding strategies, correlations measured in Euclidean distance, e.g., between local updates and historical updates, can be leveraged to establish practical coding schemes for wireless FL systems, thereby improving the efficiency of model compression. 

\subsection{Task-Oriented Model Aggregation} 
The global model aggregation involves the local model transmission from massive edge devices, followed by global model updates based on the received signal at the edge server. The conventional strategy aims at reliably recovering each transmitted message and fails to account for the model aggregation tasks and the variation of device participation across learning rounds. To scale up the model transmission, a task-oriented transmission strategy with integrated communication and computation will be presented, including task-oriented over-the-air computation (AirComp) for model aggregation and learning-aware resource rationing for device scheduling.

\subsubsection{Model Transmission} 
In each communication round, the edge server is required to update the global model, which is a specific aggregation function of local models. The model accuracy and robustness can be improved by appropriately designing the task-oriented model aggregation functions, e.g., weighted averaging aggregation functions to account for statistical heterogeneity, normalized averaging aggregation functions to avoid the objective inconsistency introduced by system heterogeneity, and geometric median of local models to achieve Byzantine-resilient robust aggregation. 
The conventional transmission strategies ignore the model aggregation task and simply perform the global model aggregation after the local models from all participating  devices are successfully decoded by the edge server. This however would incur high communication expense and prolong the training duration, as the local models are usually high-dimensional and the number of edge devices is typically large. By observing that the edge server is only required to receive a specific function of the local models instead of each local model, AirComp, as a task-oriented multiple access scheme, enables ultra-fast model aggregation \cite{yang2020over}. 
According to the principle of AirComp, all participating edge devices simultaneously transmit their local models over the same radio channel, while the edge server, via leveraging the waveform superposition property, directly receives a noisy version of the specific model aggregation function of local models. 
Due to the channel fading and receiver noise, there exists an inevitable distortion between the received global model and the error-free counterpart, the level of which determines the learning performance.
To be compatible with digital communication systems, coded AirComp can be developed based on algebraic network information theory.

\subsubsection{Device Scheduling} 
Because of limited radio spectrum resources and a large number of participating devices, device scheduling is an indispensable component of wireless FL to address the communication bottleneck and to achieve computation and load balance. 
Device scheduling for conventional data services generally assumes that the data to be transmitted is equally important across time. This however does not hold for wireless FL with diversified performance criteria and multiple learning rounds, i.e., different learning rounds may have varying impacts towards the learning performance \cite{Shen_ComMag21}. Hence, task-oriented device scheduling needs to be developed to explicitly capture the long-term impact and local model updates on the FL performance, e.g., convergence speed, model accuracy, and privacy guarantee. Specifically, with non-independent and identically distributed (non-i.i.d.) datasets across different edge devices, it is beneficial to schedule the devices that have the most informative local updates. For devices with unbalanced datasets, diverse channel conditions, and heterogeneous computation capabilities, scheduling the devices that have larger datasets, higher channel gains, and more powerful computation capability is desirable to reduce the training time. 
Instead of maximizing the number of admitted users in conventional communication systems, wireless FL only increases the number of participating devices whenever they contribute to the learned global model.
The joint design of model aggregation and probabilistic device selection can enhance the differential privacy while guaranteeing the learning performance. 
Furthermore, developing device scheduling from the temporal perspective can achieve an optimal tradeoff between learning performance and communication efficiency, for which scheduling more devices in later learning rounds is beneficial \cite{Shen_ComMag21}. 

\begin{figure*}[t] \centering
\includegraphics[scale = 0.53]{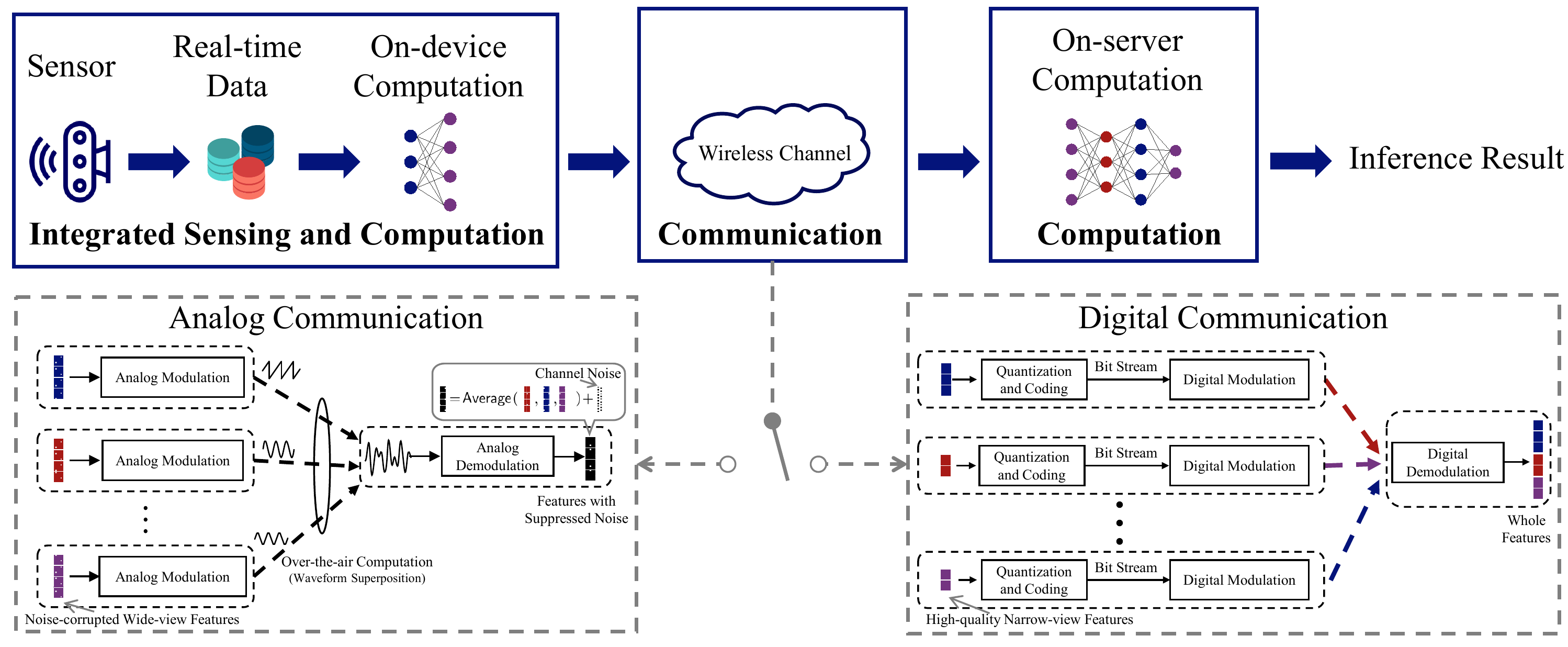}
\caption{An illustration of task-oriented edge AI inference with integrated sensing, communication, and computation.}
 \label{Fig:2}
\end{figure*}
\vspace{-2mm}

\section{Task-Oriented Communications for Edge AI Inference}
In this section, we present ``inference task"-oriented feature compression and transmission schemes.

\subsection{Vision}

Deploying well-trained AI models at the network edge facilitates the fast access of sensory data for reducing the service latency.
Edge inference, including on-device inference, on-server inference, and device-server co-inference, has great potential for empowering various edge AI applications (e.g., auto-driving, Metaverse). 
However, on-device inference achieves low accuracy due to limited computation resources, while on-server inference violates the data privacy and faces the communication bottleneck due to the transmission of high-dimensional raw data. 
To address these issues, a device-edge co-inference framework, called edge split inference, emerges as a promising solution. 
Specifically, an AI model is divided into two sub-models, where one sub-model is executed on edge devices for feature extraction and the other sub-model is deployed at the edge server for fast computation.

To provide real-time and high-quality edge inference services, novel task-oriented communication and resource allocation methods are essential. 
Performing an inference task demands useful extracted features instead of raw data, which evokes task-oriented compression. Besides, different task-relevant feature elements have diverse contributions to inference accuracy, which motivates the design of task-oriented transmissions to enhance resource utilization. 
Edge split inference typically consists of three coupled processes, i.e., sensing for data acquisition, communication for feature transmission, and computation for AI model execution. The inference performance depends on the feature distortion caused by sensing noise, hostile wireless channels, and limited processing capabilities. This motivates  task-oriented integration of sensing, communication, and computation (ISCC) schemes to unleash the full potential for the provision of intelligent services, as shown in Fig. \ref{Fig:2}. 
Compared to separate design, where the three modules are individually designed, ISCC achieves higher resource utilization by coordinating and sharing resources among different modules. Besides, in separate design, the design criteria of different modules are different and may not be consistent with the ultimate goal of inference tasks. 
In contrast, ISCC directly adopts inference performance as the design criterion, guided by which, the inference task can be effectively completed via the cooperation of sensing, communication, and computation.

\subsection{Task-Oriented Feature Compression}

Edge split inference targets at extracting and transmitting informative but compact features, which have sufficient information to achieve high accuracy. To achieve this goal, IB is a prominent approach to characterize this rate-distortion tradeoff \cite{shao2021learning}. 
This is achieved by maximizing the mutual information between the inference result and the extracted features, while minimizing the mutual information between the input data and the features. However, constructing the optimal mapping functions (i.e., from raw data to feature extraction and from extracted features to inference result) is a variational optimization problem with functional variables, which is intractable. To address this issue, DNN can be utilized to parameterize the mapping function, followed by learning the parameters from data. 

Apart from data compression, another essential goal of communication system design is to achieve high transmission rates over noisy channels. By combining the goals of relevant information extraction and transmission rate maximization, we propose a new principle, termed robust information bottleneck (RIB), which improves the IB rule by maximizing the mutual information between the transmitted representation and the received representation. RIB characterizes the tradeoff between the informativeness of the encoded representations and the robustness to information distortion in the received representations. It addresses the tension between improving the robustness and keeping sufficient relevant information for downstream inference tasks without incurring extra communication overhead. Due to the computational intractability of mutual information, a tractable variational upper bound of the RIB objective is derived by utilizing the variational approximation technique. The variational distribution is further parameterized by the learning-based inference model at the receiver and optimized via end-to-end training.

\subsection{Task-Oriented Feature Transmission}

In edge split inference, the sensing and communication processes compete for radio resources to suppress sensing noise and enhance communication rate, respectively. 
Meanwhile, the on-device computation result, i.e.,  number of extracted feature elements and quantization bits of each element, determines the communication load. 
Hence, the achievable inference accuracy depends on the received number of feature elements and their distortion caused by sensing noise and quantization errors. 
To fully exploit the scarce network resources, it is desirable to design task-oriented ISCC for edge split inference. 
In \cite{wen2022task}, multiple devices, each of which senses the source in a narrow view by focusing in a single angle to obtain high-quality sensory data, transmit quantized feature subsets extracted from the real-time sensory data to the edge server, where all feature subsets are further cascaded to form the whole feature vector to complete the downstream inference task. The influence of the sensing noise, the communication capacity, and the quality of the computational results on the inference accuracy is characterized, followed by developing an ISCC resource management scheme to maximize the inference accuracy. 

Although the above scheme is effective for multi-device narrow-view sensing, collecting multiple wide-view sensory data for enhancing the inference accuracy remains unexplored, where each device senses the source in a wide view for wide angle object detection and the obtained sensory data are noise-corrupted. 
A common approach to suppress the sensing noise is to average out it by aggregating all local feature vectors. 
To overcome the communication bottleneck arising from simultaneously accessing multiple devices, AirComp is an effective solution. 
The conventional design criterion, e.g., minimum mean squared error (MMSE), may not be effective, where all feature elements are regarded to be equally important. 
In fact, different feature elements have heterogeneous contributions to the inference accuracy. 
An effective method is to follow the task-oriented principle and adopt the inference accuracy as the design criterion. Under this design principle, the inference accuracy can be enhanced by developing a task-oriented joint transmit precoding and receive beamforming approach. 

However, the design target of inference tasks, i.e., instantaneous inference accuracy, depends on  various factors, e.g., types and capability of AI models, and the feature distribution. 
More importantly, inference accuracy is unknown before feeding the feature vector into the AI model. 
Thus, inference accuracy generally has no mathematical model, which imposes a critical challenge on the implementation of the above two schemes. 
To handle this issue, we adopt an approximate but tractable mechanism, called discriminant gain, for classification tasks. 
Based on the assumptions that all feature vectors are generated from the same source with Gaussian mixture model and that each Gaussian component corresponds to one class, discriminant gain measures the Kullback-Leibler (KL) divergence between arbitrary two distributions. 
With a larger discriminant gain, different classes are easier to be separated in the feature space, leading to a greater inference accuracy. 
Furthermore, discriminant gain only depends on the class distributions in the feature space and provides a theoretical maximal achievable inference accuracy, which is irrelevant of the underlying AI models.

\begin{figure*}[t] \centering
\includegraphics[scale = 0.5]{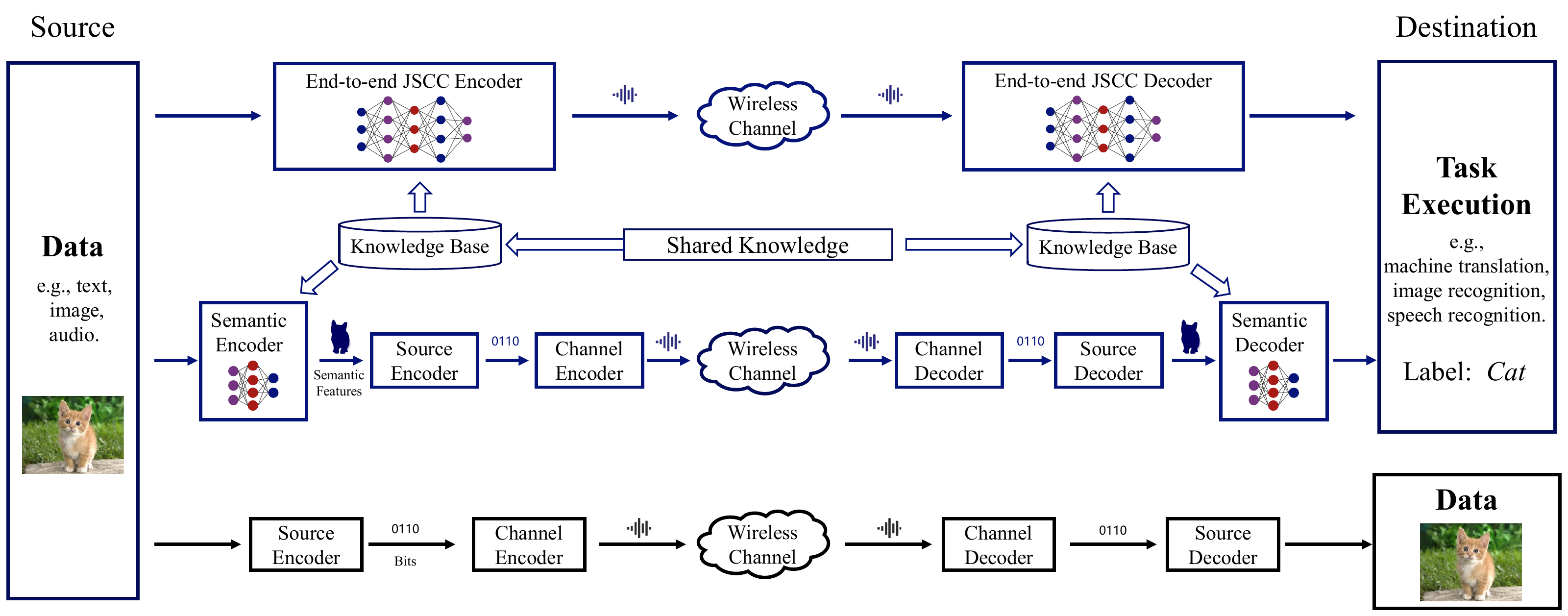}
\caption{A comparison between conventional and task-oriented semantic communication systems.}
 \label{Fig:3}
\end{figure*}

\vspace{-2mm}

\section{Task-Oriented Communications for Semantic Transmission}
In this section, we present the motivation and frameworks for task-oriented semantic communications, followed by introducing JSCC with digital modulations.

\vspace{-2mm}
\subsection{Vision}

It is envisioned that the existing communication technology may soon reach a resource bottleneck due to the stringent requirements, such as ultra-broadband, super-massive access, ultra-reliability, and low latency. Current communication systems are built on Shannon’s information theory, which aims at reliably communicating the bit sequence and ignores the meaning of the transmitted messages. However, many applications are task-oriented and the task-relevant information is sufficient to support downstream tasks. 
For instance, instead of transferring the whole sensing data, safety-related information such as obstacles or pedestrians could be detected and sent to self-driving cars to reduce communication latency while keeping safety. 
Semantic communication is a novel paradigm that integrates the meaning of information and the goal of the task into data sensing, procession, and transmission \cite{qin2021semantic}. 
By selectively sensing task-relevant data, and extracting and transmitting only task-relevant information with the assistance of some knowledge base, semantic communication can improve communication efficiency and achieve low end-to-end latency.

Fig. \ref{Fig:3} shows two types of architectures for implementing semantic communication. In the first type, the task-relevant information hidden in the source data is extracted and compressed via a semantic source encoder, followed by digital communications which convert the semantic information into bits and send them via digital modulations. Since digital communications are widely used in modern communication systems, the first type is easy to implement and only needs to focus on designing semantic source coding.  The second type applies an end-to-end JSCC strategy that directly maps the source data to channel symbols without the binary interface. 
As Shannon’s separate source-channel coding optimality holds under the assumptions of long block-length and bit-level transmission, end-to-end JSCC outperforms separate design in low-latency and semantic-level transmission by jointly extracting, compressing, and delivering the semantic information. 
This, however, may cause an incompatibility issue in modern digital communication systems that ignore the semantic meaning of data and use separate source-channel coding strategies.

Despite various endeavors in semantic communications, the following fundamental questions are still obscure: 1) How to measure the amount of semantic information hidden in the source and transmitted over channels; 2) How to optimally extract and compress semantic information; 3) How to build and exploit knowledge bases at the transmitter and receiver to improve the compression rate and communication efficiency; 4) How to optimally deliver the semantic information in an end-to-end goal-oriented manner; and 5) How to implement semantic communications in practical scenarios \cite{zhang2022toward}. In the following, we discuss semantic communications from the perspectives of semantic source coding and JSCC, and propose a unified framework to characterize semantic source coding and easy-implement semantic communication systems. 

 \vspace{-2mm}
\subsection{Semantic Source Coding}

Given the task objectives and knowledge base, semantic source coding extracts the most relevant and compact information and compresses the extracted information, which reduce storage cost and communication overhead. Empowered by deep learning (DL) technologies (e.g., transformer, generative adversarial network), DL-based techniques can efficiently extract and compress semantic information. However, the weights of DL models are learned from a training phase, whose results do not come with theoretic and optimality guarantees. Besides, semantic source coding could be analyzed with information-theoretic tools such as rate-distortion when knowing the distortion measures and statistic joint distributions on the source and semantic information. Unfortunately, since the semantic information is related to the task objective and mostly hidden in the source data, the statistic joint distributions are hard to obtain, thereby posing challenges on implementing the information-theoretic coding techniques. To embrace the merits of both DL and information-theoretic techniques, information-theoretic tools should guide the design and analysis of the DL-based semantic source coding.  

Semantic communications typically involve diversified task objectives, for which we shall propose an information-theoretic “rate-distortion-perception-classification” function that extracts and compresses the task-oriented information under data, distribution, and classification distortions. The rate-distortion-perception-classification function provides a unified theoretical tool to characterize the rate of semantic compression given various distortions of semantic communications. This includes the classical rate-distortion function with positive infinite perception and classification distortions, and the information bottleneck principle with relaxed data and perception distortion constraints. We then transform the rate-distortion-perception-classification function into a loss function and implement the semantic source coding with DNN, which requires no statistical joint distribution between the source data and semantic information. Specifically, we use deep learning networks to compress the source data, and design a loss function similar to the information bottleneck rule, but with two additional loss terms. One is the perception distortion (e.g., KL-divergence between distributions of original data and generate data), and the other is the classification distortion (e.g., the entropy of the label conditional on the compressed data). By changing Lagrange multipliers, various trade-offs among the data distortion, distribution distortion, and classification distortions can be achieved.

\subsection{Joint Source-Channel Coding}

The goal of semantic communications is to recover a subset of semantic information for tasks, rather than solely estimating the input source data. 
By jointly extracting, compressing, and delivering the semantic information, the deep learning-based joint source-channel coding (DL-JSCC) can significantly reduce communication overhead, and achieve higher communication reliability than the traditional separate source and channel coding approach. The gain of DL-JSCC can be explained in three-fold. 
First, DL-JSCC learns the most relevant information to the task goal and therefore saves the communication cost. Second, in practice, the source data itself could have space correlations (e.g., image data), temporal correlations (e.g., speech signals), context correlations (e.g., language text), etc. These correlations can be exploited by DL-JSCC to outperform the separate source and channel coding whose optimality holds for i.i.d source data. 
Third, in the training phase, the network uses a training dataset to learn the hidden task-oriented information and structure of the source data. This information can be viewed as side information or knowledge base, and can be further exploited to improve compression efficiency. However, most DL-JSCC works focus on achieving the high performance of DL-JSCC, while ignoring compatibility with current digital communication systems. It becomes critical to consider both compatibility and performance of DL-JSCC for implementing semantic communication systems. 

We propose a DL-JSCC framework based on discrete representations encoding that owns the properties of high performance and good compatibility \cite{Wen_arXiv22}. Discrete representation encoding first uses the training data to learn a discrete task-related codebook that is dependent upon the dataset, and then maps each input data into a subset of codewords and outputs the indices of codewords. The amount of information carried by the codebook is intrinsically limited by the cardinality of discrete alphabets. Due to the dependence, the task-related codebook could serve as side information to further compress the data compared to the independent codebook in conventional communication systems. To perform JSCC, the channel noise is added to the discrete representation during the training, which makes the task-related codebook not only carry semantic information, but also be entitled to redundancy to combat transmission uncertainty. Subsequently, the indices of messages (semantic messages) can be transferred using the conventional channel coding and modulation. By doing so, this method not only achieves performance gain of the joint source-channel coding, but also can be implemented in the current communication systems.  

\section{Conclusion}

Task-oriented communication is a new paradigm for wireless communication systems which goes beyond bit-level transmission to focus on intelligent task experiences. Such paradigm for disruptive technologies for information processing, signal transmission, service provision, and network resource management. The challenges for developing task-oriented communication ecosystems are multidisciplinary spanning information theory, communication theory, learning theory and domain applications. The viewpoints of information structures and communication goals provide a new way of investigating task-oriented communication theory and systems. Although only typical examples were presented, we hope that this article will motivate more exciting theories, methodologies, and applications for promoting the evolution of wireless communications.

\bibliography{reference}
\bibliographystyle{ieeetr}

\vspace{3mm}
{

{\bf{Yuanming Shi}} [S’13-M’15-SM’20] (shiym@shanghaitech.edu.cn) received the B.S. degree from Tsinghua University and the Ph.D. degree from The Hong Kong University of Science and Technology. He is currently a tenured Associate Professor at the School of Information Science and Technology, ShanghaiTech University. 
%He was a recipient of the 2016 IEEE Marconi Prize Paper Award in Wireless Communications, the 2016 Young Author Best Paper Award by the IEEE Signal Processing Society, and the 2021 IEEE ComSoc Asia-Pacific Outstanding Young Researcher Award.

{\bf{Yong Zhou}} [S'13-M'16-SM'22] (zhouyong@shanghaitech.edu.cn) received the Ph.D. degree from the University of Waterloo in 2015. From Nov. 2015 to Jan. 2018, he worked as a postdoctoral research fellow in the Department of Electrical and Computer Engineering, The University of British Columbia. He is currently an Assistant Professor in the School of Information Science and Technology, ShanghaiTech University.  

{\bf{Dingzhu Wen}} [S'19-M'21] (wendzh@shanghaitech.edu.cn) received Bachelor degree and Master degree from Zhejiang University in 2014 and 2017, respectively, and received Ph. D. degree from The University of Hong Kong in 2021. Currently, he is an Assistant Professor at School of Information Science and Technology, ShanghaiTech University. 

{\bf{Youlong Wu}} [S'11-M'14] (wuyl1@shanghaitech.edu.cn) received his Ph.D. degree at Telecom ParisTech, France, in 2014. He is now an Assistant Professor in the School of Information Science and Technology at ShanghaiTech University.

 {\bf{Chunxiao Jiang}} [S’09-M’13-SM’15] (jchx@tsinghua.edu.cn) is an associate professor in the School of Information Science and Technology, Tsinghua University. He received a B.S. degree in information engineering from Beihang University, Beijing, in 2008, and a Ph.D. degree in electronic engineering from Tsinghua University, Beijing, in 2013, both with the highest honors. %His research interests include application of game theory, optimization, and statistical theories to communication, networking, and resource allocation problems, in particular space networks and heterogeneous networks. 

 {{\bf{Khaled B. Letaief}} [S'85-M'86-SM'97-F'03] (eekhaled@ust.hk) received his Ph.D. degree from Purdue University. He has been with HKUST since 1993, where he was Acting Provost and Dean of Engineering, and is now a Chair Professor and the New Bright Professor of Engineering. 
From 2015 to 2018, he joined HBKU in Qatar as Provost. He is an ISI Highly
Cited Researcher and a recipient of many distinguished awards. He has served
in many IEEE leadership positions including ComSoc President,
Vice-President for Technical Activities, and Vice-President for Conferences. 
He is a Member of US National Academy of Engineering.

}

\end{document}